\documentclass{article}
\usepackage[utf8]{inputenc}
\usepackage[margin=1in]{geometry}

\usepackage{comment}
\usepackage[dvipsnames]{xcolor}
\usepackage{amsfonts}
\usepackage{graphicx}
\usepackage{setspace}
\usepackage{amsmath}
\usepackage{amssymb}
\usepackage{enumerate}
\usepackage{natbib}
\usepackage[section]{placeins}
\usepackage{subcaption}
\usepackage{dsfont}
\usepackage{babel}
\usepackage{hyperref}
\usepackage{authblk}
\usepackage{bbm}
\usepackage[export]{adjustbox} 

\makeatletter
\AtBeginDocument{%
  \expandafter\renewcommand\expandafter\subsection\expandafter{%
    \expandafter\@fb@secFB\subsection
  }%
}
\makeatother

\newcommand{\E}{\mathbb{E}}
\newcommand{\V}{\mathbb{V}}
\newcommand{\Cov}{\text{Cov}}
\newcommand\indep{\perp\!\!\!\perp}

\newcommand{\X}{\mathbf{X}}

\usepackage[toc,page,header]{appendix}
\usepackage{minitoc}

\begin{document}

\doparttoc 
\faketableofcontents 

\title{Demystifying and avoiding the OLS ``weighting problem'': Unmodeled heterogeneity and straightforward solutions}

\author[1]{Tanvi Shinkre}
\author[1,2]{Chad Hazlett}
\affil[1]{Department of Statistics and Data Science, UCLA}
\affil[2]{Department of Political Science, UCLA}

\date{March 2025}

\maketitle

\begin{abstract}
Researchers frequently estimate treatment effects by regressing outcomes (Y) on treatment (D) and covariates (X). Even without unobserved confounding, the coefficient on D yields a conditional-variance-weighted average of strata-wise effects, not the average treatment effect. Scholars have proposed characterizing the severity of these weights, evaluating resulting biases, or changing investigators' target estimand to the conditional-variance-weighted effect. We aim to demystify these weights, clarifying how they arise, what they represent, and how to avoid them. Specifically, these weights reflect misspecification bias from unmodeled treatment-effect heterogeneity. Rather than diagnosing or tolerating them, we recommend avoiding the issue altogether, by relaxing the standard regression assumption of “single linearity” to one of “separate linearity” (of each potential outcome in the covariates), accommodating heterogeneity. Numerous methods—including regression imputation (g-computation), interacted regression, and mean balancing weights—satisfy this assumption. In many settings, the efficiency cost to avoiding this weighting problem altogether will be modest and worthwhile.
\end{abstract}

\clearpage

\doublespacing

\section{Introduction}
When estimating the effect of a treatment on an outcome of interest by adjusting for covariates ($X$), researchers typically hope to interpret their result as a well-defined causal quantity, such as the average effect over strata such as the average treatment effect (ATE). Despite the introduction of many more flexible estimation procedures, regression adjustment remains the ``workhorse approach'' when adjusting for covariates to make causal claims in many disciplines \citep{aronow_does_2016}.\footnote{In political science, for example, \cite{keele_adjusting_2010} found that regression adjustment was used in analyzing 95\% of experiments in American Political Science Review, 95\% in Journal of Politics, and 74\% in American Journal of Political Science. The ubiquity of regression adjustment approaches in practice is echoed by authors in economics and other disciplines as well, such as \cite{angrist_mostly_2009,humphreys_bounds_2009, chattopadhyay2023implied}. This is perhaps due to familiarity, ease of use, the efficiency of OLS and its suitably good performance in many contexts \citep{hoffmann_2023, green_analyzing_2011, kang_demystifying_2007}, and well-established uncertainty estimation considerations and tools.} For a binary treatment ($D$) and outcome ($Y$), a simple bivariate regression of $Y$ on $D$ gives the difference in means estimate (the mean of $Y$ when $D=1$ minus the mean of $Y$ when $D=0$). However, this equivalence breaks down when covariates $X$ are included in the regression (as in $Y \sim D + X$), if the treatment effect may vary across values of $X$. Instead, the  coefficient produced by this regression represents a weighted average of strata-specific difference in means estimates, where the weights are larger for strata with probability of treatment closer to 50\%  \citep{angrist_estimating_1998, angrist_mostly_2009}. Such a weighted average is not typically of direct interest, and under severe enough heterogeneity in treatment effects, this can lead to widely incorrect substantive conclusions, as demonstrated below.

An ongoing literature regards these weights as a nuisance to be coped with or incorporated into analysis. Accordingly, authors have proposed diagnostics that index the potential for bias due to these weights or tools to aid interpretation given these weights \citep{aronow_does_2016, sloczynski_interpreting_2022, chattopadhyay_lmw_2023, chattopadhyay2023implied, kline2011}, and have analyzed the behavior of proposed regression estimators in different settings in light of these weights (e.g. \citealp{chernozhukov2013average}).

In this article, we seek to demystify these weights,  offering a simple derivation and conceptual clarification of why they arise, what they represent, and how to avoid them. We offer two main theoretical clarifications. First, under heterogeneous treatment effects (and varying probability of treatment), the linear regression of $Y$ on $D$ and $X$ will be misspecified (e.g. \citealp{rubin1979, imbens_recent_2009, imbens2015}). Re-expressing the regression coefficient as a weighted average of strata-wise effects simply offers a way to see how this misspecification causes the regression coefficient to diverge from the ATE. Such weights emerge, as we show, simply by relying on the Frisch-Waugh-Lowell (FWL) theorem \citep{frisch-waugh-1933,lovell-1963} to first construct the unit-level (not strata-level) weights. We can then group these weights to form a new expression of the strata-wise weights. Our expression does not rely on the usual assumption (as in \cite{angrist_estimating_1998} and later sources) that $D$ is linear in $X$, but reduces to that expression in the special case where $D$ is linear in $X$. This result corroborates the recent independent contribution of \cite{hahn2023properties}. 

Second, we clarify the assumption under which approaches can avoid these weights, which determines what (existing) estimators avoid this problem. 
Rather than attempting to diagnose, interpret, or otherwise cope with the undesired affect of these weights on interpretation, they can be avoided by any estimator that works under the \textit{separate linearity} assumption, meaning each potential outcome $Y(d)$ is assumed to be linear in $X$, as opposed to the \textit{single linearity} assumption that required $Y$ itself to be linear in $X$ and $D$.
Equivalent assumptions have been stated in analyses by \cite{hahn2023properties, sloczynski_interpreting_2022, kline2011} and \cite{imbens_recent_2009}. 
Fortunately, many well-known estimation approaches can be justified under the separate linearity framework including (i) regression imputation/g-computation/T-learner/multi-regression \citep{peters1941method, belson1956technique, robins_new_1986, kunzel_metalearners_2019, chattopadhyay2023implied}; (ii) regression of $Y$ on $D$ and $X$ and their interaction \citep{lin_agnostic_2013}, and (iii) balancing/calibration weights that achieve mean balance on $X$ in the treated and control groups (e.g. \citealp{hainmueller_entropy_2012}). Approaches (i) and (ii) are shown to be identical in the context of OLS models without regularization. Mean balancing weights (iii), when constructed to target the ATE, can be justified by the same assumption of separate linearity, but uses a different estimation approach and produces different point estimates. All three of these strategies produce unbiased ATE estimates under separate linearity, without the ``weighting problem'' suffered under the single linearity regression. These benefits do come at an efficiency cost, though it is low in settings with few covariates relative to sample size. 

These considerations apply broadly to observational research in which an assumption of no unobserved confounding is relied upon. However, they can also apply to the analysis of even randomized experiments, concurring with \cite{lin_agnostic_2013}. In particular, we illustrate these implications and recommendations for block-randomized experiments.

\section{Background: OLS weights}

\subsection{Setting and notation}

We consider settings where we are interested in estimating the average effect of binary treatment $D$ on outcome $Y$, while accounting for confounder $X$. The average treatment effect is defined as 
\begin{align}
    \tau_{ATE} &= \E[Y_i(1) - Y_i(0)]
\end{align}
where $Y_i(1)$ and $Y_i(0)$ denote the potential outcomes under treatment and control, respectively \citep{neyman1923applications, rubin1974estimating}. In order to estimate this average treatment effect using observed data, we require an absence of unobserved confounders, concisely expressed as the conditional ignorability assumption,
\begin{equation}
\label{cond-ig}
    D_i \indep \{Y_i(1), Y_i(0)\} | X_i
\end{equation}

For some discrete variables $X$ that satisfy conditional ignorability, and given consistency of the potential outcomes, the ATE can be identified according to:
\begin{align}
\tau_{ATE} &= \sum_{x \in \mathcal{X}} \E[Y(1)-Y(0)|X=x]P(X=x) \\ \nonumber
&= \sum_{x \in \mathcal{X}} \left(\E[Y|D=1,X=x]-\E[Y|D=0,X=x]\right) P(X=x) \nonumber \\
     &= \sum_x DIM_x P(X=x) 
\end{align}

where we suppress the $i$ subscripts for legibility and $DIM_x$ is the estimand for the difference in means in the stratum where $X=x$. In a given sample, this is approximated using the analog estimator,\footnote{This quantity could also properly be labeled as the (estimate of) the sample average treatment effect (SATE) rather than the ATE. However, we maintain the ATE notation for simplicity as its expectation over-samples is still the ATE, presuming the sample in question is a probability sample from the population of interest.}
\begin{equation}
   \hat{\tau}_{ATE} = \sum_x \widehat{DIM}_x \hat{P}(X=x) \label{avedim}
\end{equation}

 The term $\hat{P}(X=x)$ gives the empirical proportion of units in each stratum in the sample, and can be thought of as the ``natural'' strata-wise weights, as they  marginalize over the stratum according to the fraction of units falling in that stratum. It would be natural to form an analog estimator for this through sub-classification/stratification, simply computing the difference in means in each stratum of $X$ and compiling them per expression~\ref{avedim}.
However, suppose we instead attempt to estimate the treatment effect by fitting a regression according to the model
\begin{equation}
    Y = \beta_0 + \beta_1 X + \tau_{reg} D + \epsilon
\end{equation}
 
While regression-based estimation of treatment effects has been  widely used in practice across disciplines for decades, as \cite{angrist_estimating_1998} and many scholars since then have emphasized, $\hat{\tau}_{reg}$ do not in general yield $\hat{\tau}_{ATE}$, even when conditional ignorability holds. Rather, it can be understood as a version of Expression~\ref{avedim} but with weights on each $\widehat{DIM}_x$ that differ from $\hat{P}(X=x)$. We now turn to a ground-up analysis of these strata-wise weights intended to demystify them and to recognize them as the natural consequence of misspecification of the linear model under heterogeneous effects.

\subsection{The unit-level weighting representation of OLS}

While our plan is to (re-)derive strata-wise weights on the $\widehat{DIM}_x$ components, we start with ``unit-level weights'', which arise simply because the regression coefficient is a weighted sum of $Y_i$ values. Specifically, let $\hat{d}(X_i) = \hat{p}(D_i = 1 | X_i) =  \X_i \hat{\beta}$ where $\X_i = \begin{bmatrix} 1 & X_i \end{bmatrix}$,  and $\hat{\beta}$ is the estimated coefficients from a linear regression of $D$ on $X$. The unit-wise weighting formula that corresponds to the OLS estimate is defined according to

\begin{equation}\label{indivweightform}
    \hat{\tau}_{reg} = \sum_i w_i Y_i
\end{equation}

for some weights $w_i$. By the FWL theorem \citep{frisch-waugh-1933, lovell-1963},
\begin{align}
    \hat{\tau}_{reg} =& \frac{
    \widehat{\Cov}(Y_i, D_i^{\perp X})}{\widehat{\V}(D_i^{\perp X})} 
    = \frac{\sum_i Y_i ( D_i - \hat{d}(X_i))}{\sum_i (D_i - \hat{d}(X_i))^2} \\
\end{align}

\noindent which already takes the form of Equation~\ref{indivweightform}, with weights given by  
\begin{align}\label{eq.indivweights}
    w_i =& \frac{D_i - \hat{d}(X_i)}{\sum_i (D_i - \hat{d}(X_i))^2} 
\end{align}

\noindent \textit{Connection to propensity score.} Though these individual-level weights are an intermediate step in our analysis, we give some consideration to their form here. First, as the denominator is a scalar, the behavior of these weights is revealed through the numerator, $D_i - \hat{d}(X_i)$. This is simply $1 - \hat{d}(X_i)$ for treated units and $\hat{d}(X_i)$ for control units. Compare this to inverse-propensity score weights (IPW), which are proportional to $1/\hat{d}(X_i)$ for treated units and $1/(1-\hat{d}(X_i))$ for control units. Like the IPW weights, these unit-level weights in Expression~\ref{eq.indivweights} place greater weight on units when their treatment status is more ``surprising'' given the covariate values. Unlike IPW, the linear regression weights do not require constructing a ratio that has a denominator that can become close to zero, which can create explosive weights under IPW.\footnote{Equivalent unit-level weights are explored in  depth by \cite{chattopadhyay2023implied} and in \cite{Chattopadhyay2024Causation}, where the authors usefully employ these weights to analogize how OLS implicitly compares (weighted) mean outcomes for the treated to (weighted) mean outcomes for the controls, sharpening the analogy to the difference-in-means estimator one might apply in a randomized experiment and discussing implications for qualities such as effective balance and sample size.} This similarity of the unit-level weights to the propensity score approach is explored in detail in \cite{kline2011}.
\bigskip

\noindent \textit{Negative individual-level weights.} The weights $w_i$ are constructed to be both positive and negative because Equation~\ref{indivweightform} is structured as a single weighted sum/average. An isomorphic but useful way of signing the weights is to instead seek the weights that would appear in a ``weighted difference in means'' estimator, 
\begin{equation}\label{tildew}
\hat{\tau}_{wdim} = \sum_{i:D=1} \tilde{w}_i Y_i - \sum_{i:D=0} \tilde{w}_i Y_i
\end{equation}

\noindent where $\tilde{w}_i = D_i w_i + (1-D_i)(-w_i)$, which simply changes the sign for control units in order to accommodate the subtraction rather than the summation in the form of Expression~\ref{indivweightform}. 

The weights $\tilde{w}$ signed as in Expression~\ref{tildew} will often be positive, and any positive weight may be naturally interpreted as the relative contribution of each observation. However, they can be negative as well. \cite{Chattopadhyay2024Causation} note that negative weights ``translate into forming effect estimates that may lie outside the support or range of the observed data'',  or that they ``[leave] room for biases due to extrapolation from an incorrectly specified model.''  \cite{borusyak2024negative} also discuss negative weights, emphasizing that under randomization, these weights will all be non-negative, and so do not pose a problem for the resulting estimand. 


We note that the meaning of negative individual weights is closely tied to estimates one would obtain from a linear probability model regressing the treatment indicators on $X$, even though that model is not run.
Specifically, weights will be negative for a treated unit when $\hat{d}(X_i)>1$, and for control units when $\hat{d}(X_i)<0$. If a treated unit lies at an $X$ position ``so unique to treated units'' that a linearly-fitted model produces a predicted value greater than 1 there, it will have a negative weight. This can be thought of as indicating that a treated unit is ``more like treated units than any control units'', so that it cannot be well compared to control units. The symmetric argument holds for control units with a negative value.  

In this way, negative individual level weights do indicate extreme non-overlap, but in a very peculiar, model dependent sense and not in any non-parametric way that corresponds to broader notions of common support. These weights are also not a sure diagnostic for poor overlap or extreme counterfactuals: one could easily imagine a situation where all units above a certain value of $X$ are treated and all those below another value of $X$ are controls, indicating poor overlap. However, there can be many (or all) units outside the area of common support but with positive weights, because the fitted linear probability model for $D$ given $X$ does not exceed $1$ (for treated units) or drop below $0$ (for control units). Thus, we emphasize that negative weights can provide a warning that ``some units look too much like treated units to have valid comparisons'' (and likewise for controls), but a positive weight is not a guarantee of meaningful comparability in the sense of finding units of the opposite treatment value nearby in the covariate space.

\subsection{From individual to strata-wise weights}\label{subsec.strataweights}

With the individual-level weights in hand, we can now construct the strata-level weights of interest to our analysis. These weights are interesting because they allow us to see how the regression coefficient compares to ATE as a combination of  average treatment effects by stratum as in Equation~\ref{avedim}. To do so we consider the case of discrete $X$ and the strata-wise weights, meaning those that can be expressed as,

\begin{equation}
    \hat{\tau}_{reg} = \sum w_x \hat{\tau}_x
\end{equation} 

\noindent where $w_x$ is the weight for the strata in which $X=x$ and $\hat{\tau}_x = \hat{\E}[Y_i(1) - Y_i(0) | X_i = x]$ is the conditional average treatment effect for subgroup $X_i = x$. The resulting $\hat{\tau}_{reg}$ correspond to the ATE only when $w_x = P(X=x)$. The best known expression for such strata-wise weights of OLS arises from \cite{angrist_estimating_1998}, 
\begin{equation}
\label{varweight-equation}
    \hat{\tau}_{reg} = \frac{\sum_x \hat{\tau}_x \hat{d}(X_i) (1 - \hat{d}(X_i))\hat{P}(X_i=x)}{\sum_x \hat{d}(X_i)(1 - \hat{d}(X_i))\hat{P}(X_i=x)}
\end{equation}

\noindent This form shows that regression weights strata-wise $\widehat{DIM}_x$ components not according to $\hat{P}(X_i=x)$ as required to represent the ATE, but proportionally to $\hat{d}(X_i) (1 - \hat{d}(X_i))\hat{P}(X_i=x)$. Such a regression puts the most weight on strata in which the conditional variance of treatment status is largest, i.e. when $\hat{d}(X_i)$ is nearest to 50\%. An intuition for this notes that OLS seeks to minimize squared error, and the opportunity to learn the most from strata with middling probabilities of treatment \citep{angrist_mostly_2009}.  Absent treatment effect heterogeneity, this is unproblematic. However, if there are high levels of effect heterogeneity in our data, then depending on how they correspond to strata of $X$ and the probability of treatment in those strata, these weights could move the regression coefficient far from the average treatment effect \citep{ aronow_does_2016, humphreys_bounds_2009, angrist_mostly_2009,sloczynski_interpreting_2022}. \cite{goldsmith2022contamination} also discusses similar weights in the context of estimating effects for multiple treatments at once, where this leads to ``contamination'' of effect estimates.

These weights, however, are obtained under the additional assumption that the probability of treatment is linear in $X$, and not just modeled that way. \cite{angrist_mostly_2009} satisfy this by assuming the corresponding outcome regression can be saturated in $X$.  This is often a reasonable assumption with discrete, low-dimensional $X$. Nevertheless we may wish to have a more general expression for the weights that does not rely on such an assumption, either for completeness or in service of generalizing to the case where $X$ is not discrete or is otherwise infeasible to include in a saturating form (i.e. $X$ with numerous multiple dimensions and levels). The issues that could arise when this type of assumption is not satistifed are highlighted by \cite{NBERw29709}, who similarly analyze the weighted average representation of the two stage least squares estimate for the local average treatment effect presented by \cite{angrist_mostly_2009}, and show that the equivalent linearity assumption that is invoked is often not satisfied in practice.

To obtain a more general expression for the weights, we simply organize the individual-level weights above into strata, 
\begin{align}
    \hat{\tau}_{reg} &= \sum_{i} \ w_i Y_i \\
    &= \frac{\sum_{i} Y_i (D_i - \hat{d}(X_i))}{\sum_i(D_i - \hat{d}(X_i))^2} \\
    &= \frac{\sum_x \hat{P}(X_i=x) \widehat{\E}[(D_i - \hat{d}(X_i)) Y_i | X_i=x]}{\sum_x \hat{P}(X_i=x) \widehat{\E}[(D_i - \hat{d}(X_i))^2 | X_i=x]}
\end{align}
Rearranging these terms in the form $\hat{\tau}_{reg} = \sum_x w_x \tau_x$ is not in general possible, as the expected outcome under treatment and the expected outcome under control are weighted differently:
\begin{align}
    \widehat{\E}[(D_i - \hat{d}(X_i)) Y_i | X_i = x] &= \widehat{\E}[D_i (1 - \hat{d}(X_i)) Y_i- (1 - D_i) \hat{d}(X_i) Y_i | X_i = x]  \\
    &=  (1 - \hat{d}(X_i))\widehat{\E}[D_i Y_i | X_i = x] - \hat{d}(X_i) \widehat{\E}[(1 - D_i) Y_i | X_i = x]  \\
     &=  (1 - \hat{d}(X_i)) \pi_x \widehat{\E}[Y_i | X_i = x, D_i = 1] \\ &- \hat{d}(X_i) (1 - \pi_x) \widehat{\E}[Y_i | X_i = x, D_i = 0] \nonumber
\end{align}

where $\pi_x = \hat{P}(D_i=1 | X_i = x)$ is the true, or correctly specified, probability of treatment in the sample for a given stratum. Thus the strata-wise weighting generally imposed by OLS for a given sample involves a combination of the true probability of treatment in the sample given $X$, and the probability of treatment given $X$ estimated using a linear model. This result is equivalent to the weighted representation independently developed by \cite{hahn2023properties}, who also considers a special case for when the outcome model for the control group is linear. We can also formulate our result in terms of the discrepancy, $a_x$, between the true and linearly-approximated probability of treatment given $X$, so that $\hat{d}(X_i) = \pi_x + a_x$. The general strata-wise weights corresponding to OLS are then
\begin{equation}
\label{trueweights}
    \hat{\tau}_{reg} = \frac{\sum_x \hat{P}(X_i=x) \pi_x (1 - \pi_x) \tau_x - a_x \hat{\E}[Y_i|X_i=x]}{\sum_x \hat{P}(X_i=x) (\pi_x(1 - \pi_x - a_x) + a_x(\pi_x + a_x))}
\end{equation}

Appendix \ref{derive-realweights} gives additional details. Notice that this expression is infeasible to compute when $D$ is not linear in $X$, as $\pi_x$ is unknown. However, if $a_x = 0$ (i.e. $D$ is truly linear in $X$), this representation reduces to the variance-weighted representation of the regression coefficient given by \cite{angrist_estimating_1998}. For many purposes then the \cite{angrist_estimating_1998} representation provides a clear and intuitive conception of the weights. Nevertheless, we found it necessary to have this more complete formulation in 
order to obtain the correct answer, as our simulations below show.

\subsection{From single linearity to separate linearity}

An OLS model regressing $Y$ on $D$ and $X$ alone would be correctly specified if the true conditional expectation function, $\E[Y | X, D]$ is linear in $D$ and $X$ as in $\E[Y | X, D] = \beta_0 + \beta_1 X + \beta_2D$. We call such an assumption the \textit{single linearity} assumption, because it requires a single assumption about $Y$ being linear in some terms. Note that this allows for random variation in treatment effects not correlated with $X$, but it does not accommodate treatment effect variation that is correlated with $X$. 

Accordingly, if the ATE rather than the regression coefficient is the target of inference, a natural solution is not to characterize or diagnose or bound the difference between the coefficient and the ATE, but rather to avoid the underlying misspecification problem. The strata-wise weights above merely describe the regression coefficient in other terms, and as such, elucidate the impact of misspecification for the difference between the coefficient and the ATE.  Resolving this misspecification bias is a natural solution.

Here we consider the minimal change to regression practice that an investigator otherwise comfortable with a linear model could employ to avoid this misspecification concern and the resulting weighting behavior. We first define the \textit{separate linearity} assumption, requiring that each of the potential outcomes is separately linear in $X$,
\begin{equation}
    \E[Y(0) | X] = \alpha_0 + \alpha X
\end{equation}
\begin{equation}
    \E[Y(1) | X] = \gamma_0 + \gamma X
\end{equation}
\noindent  This assumption is slightly weaker than single linearity, which effectively forces $\alpha=\gamma$. While we label this assumption for easy reference and to distinguish it from single linearity, it is not intended to be novel, and further can be understood as the (often implicit) motivation for a number of longstanding approaches described next \citep{imbens2015, kline2011, imbens_recent_2009}.  

We turn next to a variety of known and straightforward estimation approaches that are in keeping with this assumption and that wholly avoid the awkwardness of ``regression's weighting problem'' rather than employing diagnostic or interpretational aids.

\subsection{Estimation approaches suitable for separate linearity} 

\paragraph{Interactions, imputation, g-computation, and stratification.}

Fortunately, a number of existing approaches are suggested by such an assumption. One alternative estimation approach, proposed by \cite{lin_agnostic_2013} initially to mitigate bias in covariate-adjusted estimates from randomized experiments, is to use a regression model that includes an interaction term for the treatment and confounder. \cite{goldsmith2022contamination} similarly propose interacted models in the context of multi-valued treatments. 

In the binary treatment setting we consider, the treatment effect estimate, $\hat{\tau}_{interact}$, is the estimated coefficient on the treatment variable $D_i$ in the regression of $Y_i$ on $D_i$, $X_i - \bar{X}$, and $D_i(X_i - \bar{X})$. The centering of $X$ in these terms is useful for interpretation: while the fitted models with and without centering $X$ are isomorphic, centering $X$ allows the coefficient on $D$ to represent the average effect ``when $X$ is at its mean''. By linearity, this is also exactly the average marginal effect of $D$ on $Y$ taken across observations at their observed values of $X$.\footnote{This centering procedure and the resulting interpretation is complicated when one level of $X$ (or the intercept) must be dropped, as when $X$ represents block indicators or more generally group fixed effect. We describe this concern and propose a solution for this in Section~\ref{blockrandomization}.} 

Another straightforward  approach to estimate the ATE under separate linearity is to run separate regressions of $Y$ on $X$ for the treatment group and the control group, and use the results from each regression to predict the unobserved outcomes in the other group. Then, using these predicted outcomes, we can estimate the individual treatment effect for each subject, and take the average of the estimated individual treatment effects to get the ATE. This is also known as g-computation, meaning that an estimate from g-computation will also be unbiased for the ATE in settings with high levels of heterogeneity in treatment effect and probability of treatment assignment \citep{robins_new_1986, snowden_implementation_2011}. It has also been referred to even more recently as ``multi-regression'' \citep{Chattopadhyay2024Causation}, who note earlier uses of this approach as far back as \cite{peters1941method} and \cite{belson1956technique}. It is also equivalent to the Oaxaca-Blinder decomposition \citep{oaxaca1973, blinder1973}.

Specifically, let $\hat{\mu}_0(X_i)$ be the model fit to the control units and $\hat{\mu}_1(X_i)$ be the model fit to the treated units,
\begin{align}
    \hat{\tau}_{imp} &= \frac{1}{n} \left(\sum_{D_i=1} (Y_i - \hat{\mu}_0(X_i)) + \sum_{D_i = 0} (\hat{\mu}_1(X_i) - Y_i )\right) \label{imp1}\\
    &= \frac{1}{n}\left( \sum_{D_i=1} ((\hat{\mu}_1(X_i) + \epsilon_{1i}) - \hat{\mu}_0(X_i)) + \sum_{D_i = 0} (\hat{\mu}_1(X_i) - (\hat{\mu}_0(X_i) + \epsilon_{0i})) \right)\\
     &= \frac{1}{n} \left( \sum_i \hat{\mu}_1(X_i) - \hat{\mu}_0(X_i) \right) \label{imp2}
\end{align}

Because this explicitly puts weights of $1/N$ on every unit, the estimate does not suffer from the ``weighting problem".\footnote{We note the equality of Expression~\ref{imp1} to Expression~\ref{imp2} above implies that one may either (i) compare each unit's observed outcome under the realized treatment status to the modeled outcome for that unit under the opposite, or (ii) for each unit, compare the \textit{modeled} outcome under treatment to the \textit{modeled} outcome under control, without using the observed outcome. This is a result of relying on OLS for each outcome model, since the average fitted value from a given model will be precisely equal to the average observed outcome over the same group. Such a property does not hold with estimators that cannot guarantee $\overline{\hat{\epsilon}}=0$.} Further, the Lin estimate and the regression imputation estimate are easily shown to be identical in this context. Specifically, the Lin estimate models $\E[Y|D,X]$ as $\beta_0 + \beta_1 D + \beta_2 X + \beta_3 DX$, and thus implies
\begin{equation}
    \E[Y | X, D=0] = \beta_0 + \beta_2 X
\end{equation}
\begin{equation}
    \E[Y | X, D=1] = (\beta_0 + \beta_1) + (\beta_2 + \beta_3) X
\end{equation}
Meanwhile, the imputation estimate is based on two regression models, one for the untreated group and one for the treated group:
\begin{equation}
    \hat{\mu}_0(X_i) = \E[Y| X, D=0] = \alpha_0 + \alpha_1 X
\end{equation}
\begin{equation}
    \hat{\mu}_1(X_i) = \E[Y| X, D=1] = \gamma_0 + \gamma_1 X
\end{equation}
Comparing this to the above equations, we can prove $\alpha_0=\beta_0$, $\alpha_1 =\beta_2$, $\gamma_0 = \beta_0 + \beta_1$, and $\gamma_1 = \beta_2 + \beta_3$ by showing the equivalency of the minimization problems in question (see Appendix \ref{equiv} for details). Using these equivalencies, we can show that the ATE estimate from regression imputation is equivalent to the estimate from the interacted regression,
\begin{align}
    \hat{\tau}_{imp} 
    &= \frac{1}{n} \sum_{i=1}^n (\gamma_0 + \gamma_1 X_i - (\alpha_0 + \alpha_1 X_i)) \\
    &= (\gamma_0 - \alpha_0) + (\gamma_1 - \alpha_1) \bar{X} \\
    &= (\beta_0 + \beta_1 - \beta_0) + (\beta_2 + \beta_3 - \beta_2) \bar{X} \\
    &= \beta_1 = \hat{\tau}_{interact}
\end{align}
Since these two estimation methods are identical under OLS, they can be used interchangeably. One can directly show the unbiasedness of either under assumptions of consistency and conditional ignorability,

\begin{align}
    \E[\hat{\tau}_{imp}] &= \E\left[\frac{1}{N} \sum_{i=1}^N \left(\hat{\mu}_1(X_i) - \hat{\mu}_0(X_i)\right)\right] \\
    &= \E[\E[Y_i | D_i = 1, X_i]] - \E[\E[Y_i | D_i = 0, X_i]] \\
    &= \E[Y_i(1) - Y_i(0)]
\end{align}

Both are also identical to the stratification estimate when $X$ is discrete,
\begin{align}
    \hat{\tau}_{interact} = \hat{\tau}_{imp} &= \frac{1}{N} \sum_{i=1}^N \big(\hat{\mu}_1(X_i) - \hat{\mu}_0(X_i)\big) \\
    &= \sum_{x \in X} \hat{P}(X=x) \widehat{\E}[\hat{\mu}_1(X_i) - \hat{\mu}_0(X_i)| X_i = x]
\end{align}

\paragraph{Mean balancing weights.} We can also connect the separate linearity  assumption to a justification for calibration/balancing weights. Consider mean balancing weights for both the treated and control units, so that weighted average of covariates for each is equal to the overall unweighted average of covariates, 
\begin{equation}
     \sum_{i: D_i=0} w_i X_i = \sum_{i: D_i=1} w_i X_i  = \frac{1}{N}\sum_{i=1}^{N} X_i
\end{equation}

While many choices of weights can satisfy these constraints (when feasible), it is desirable to minimize some measure of their variation. In the case of  
maximum entropy weights \citep{hainmueller_entropy_2012}, this is done by maximizing the entropy, $\sum_i w_i log(w_i/q_i)$. The key idea is that by achieving equal means on $X$ in the treated and control group (both equal to the full samples mean on $X$), then any linear function of $X$---which include $Y(1)$ and $Y(0)$ if separate linearity holds--- will also have equal means in these groups. A difference in means estimator using these weights is then unbiased for the ATE, without requiring any appeal to the relationship between this weighting procedure and propensity score modeling.\footnote{See Appendix \ref{meanbal-unbiased} for proof, with similar results targeting the ATT in \cite{hazlett2020kernel}. See \cite{zhao2017entropy} for a detailed analysis of identification and double-robustness of mean balancing weights.} \cite{kline2011} discusses how the regression imputation approach can be formulated as a weighting estimator that balances the means of covariates. An interesting feature of the mean balancing approach is that while it is justified by the same assumptions as regression imputation or the interactive regression above, their estimation strategies are different. Balancing weights do not require estimating the (nuisance) coefficients of any model.   This may improve tolerance to misspecification (i.e. non-linear conditional expectations for each potential outcome) at the  cost of variance, as borne out in simulations below.

\section{Simulations}

To verify that these estimators behave as our analysis claims, we explore the performance of different estimation approaches under three different data generating processes, first with a discrete covariate $X$ and then with a continuous one.

\subsection{Discrete covariate}
\begin{figure}[htp!]
        \caption*{DGP1: $P(D|X)$ increasing in $X$; $\{Y(1), Y(0)\}$ linear in $X$}
\begin{subfigure}{0.5\textwidth}

            \begin{tabular}{@{\extracolsep{5pt}} ccccc} 
\\[-1.8ex]\hline 
\hline \\[-1.8ex] 
 & $X$ & $P(D \vert X)$ & $P(X=x)$ & $\tau_x$ \\ 
\hline \\[-1.8ex] 
1 & -$3$ & $0.100$ & $0.143$ & -$9$ \\ 
2 & -$2$ & $0.100$ & $0.143$ & -$6$ \\ 
3 & -$1$ & $0.100$ & $0.143$ & -$3$ \\ 
4 & $0$ & $0.500$ & $0.143$ & $0$ \\ 
5 & $1$ & $0.500$ & $0.143$ & $3$ \\ 
6 & $2$ & $0.500$ & $0.143$ & $6$ \\ 
7 & $3$ & $0.500$ & $0.143$ & $9$ \\ 
\hline \\[-1.8ex] 
\end{tabular} 
        \label{fig:1a}
    \end{subfigure}%
    \begin{subfigure}{0.5\textwidth}

        \includegraphics[width=0.9\textwidth, valign=m]{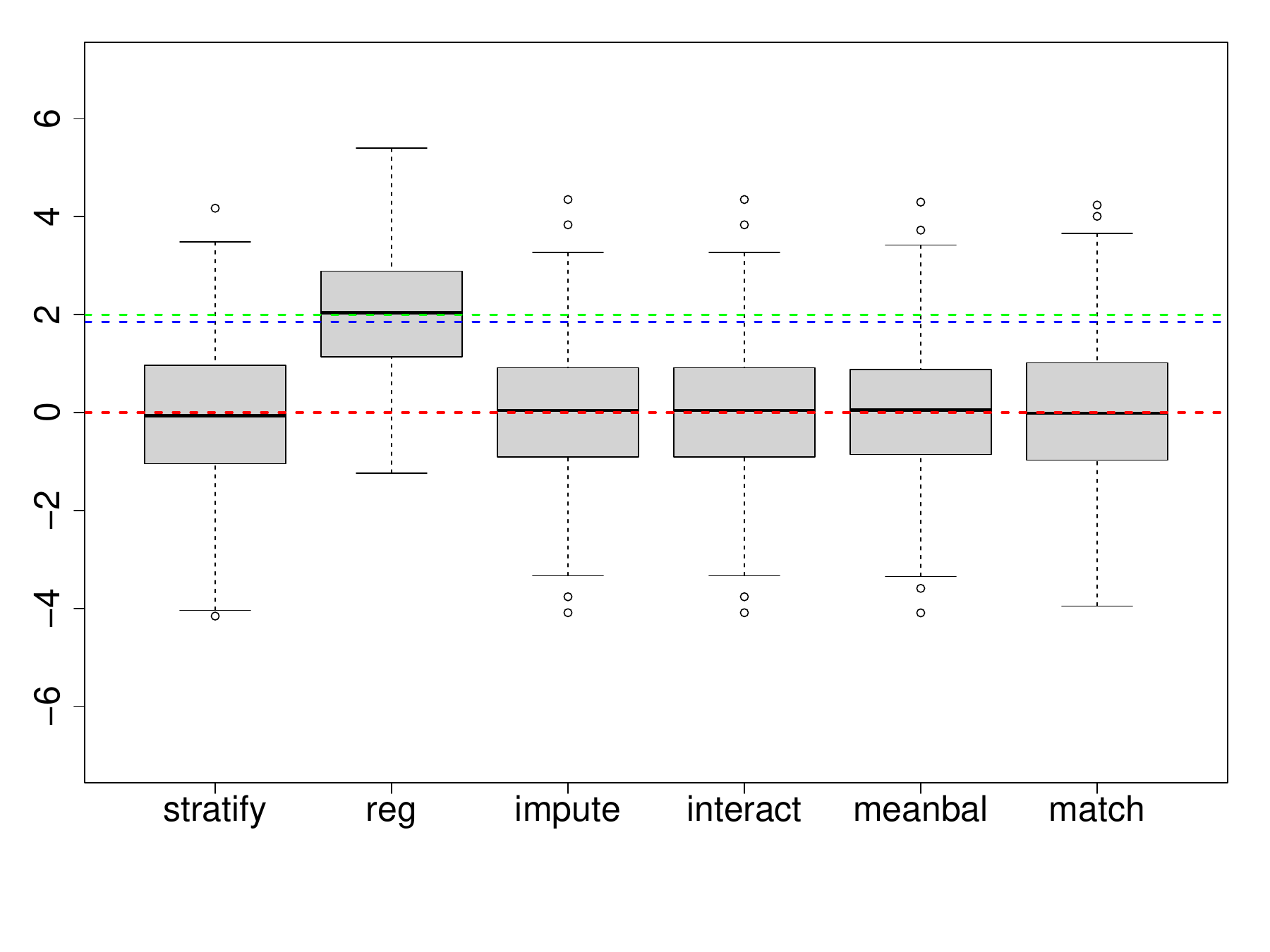} 
        \label{fig:1b}
    \end{subfigure}
        \caption*{DGP2:  $P(D|X)$ increasing linearly in $X$; $
    \{Y(1), Y(0)\}$ linear in $X$}

\begin{subfigure}{0.5\textwidth}
    
\begin{tabular}{@{\extracolsep{5pt}} ccccc} 
\\[-1.8ex]\hline 
\hline \\[-1.8ex] 
 & $X$ & $P(D \vert X)$ & $P(X=x)$ & $\tau_x$ \\ 
\hline \\[-1.8ex] 
1 & -$3$ & $0.100$ & $0.143$ & -$9$ \\ 
2 & -$2$ & $0.200$ & $0.143$ & -$6$ \\ 
3 & -$1$ & $0.300$ & $0.143$ & -$3$ \\ 
4 & $0$ & $0.400$ & $0.143$ & $0$ \\ 
5 & $1$ & $0.500$ & $0.143$ & $3$ \\ 
6 & $2$ & $0.600$ & $0.143$ & $6$ \\ 
7 & $3$ & $0.700$ & $0.143$ & $9$ \\ 
\hline \\[-1.8ex] 
\end{tabular}

    \end{subfigure}%
    \begin{subfigure}{0.5\textwidth}
       \includegraphics[width=0.9\textwidth, valign=m]{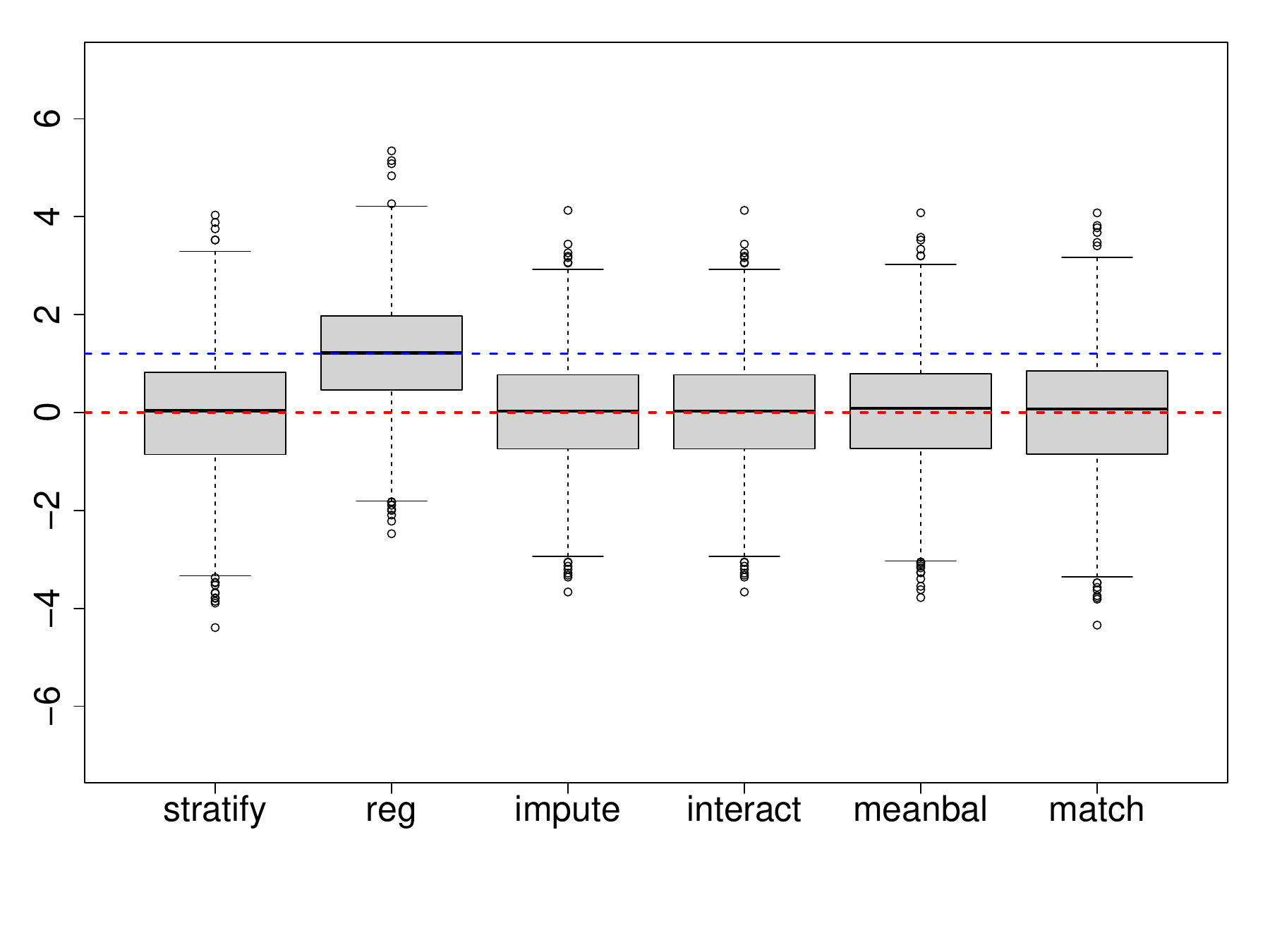} 
    \end{subfigure}

        \caption*{DGP3:  $P(D|X)$ increasing in $X$; nonlinear $Y(1), Y(0)$}
\begin{subfigure}{0.5\textwidth}
\begin{tabular}{@{\extracolsep{5pt}} ccccc} 
\\[-1.8ex]\hline 
\hline \\[-1.8ex] 
 & $X $& $P(D \vert X)$ & $P(X=x)$ & $\tau_x$ \\ 
\hline \\[-1.8ex] 
1 & -3 & $0.100$ & $0.143$ & $5$ \\ 
2 & -$2$ & $0.200$ & $0.143$ & $5$ \\ 
3 & -$1$ & $0.300$ & $0.143$ & $0$ \\ 
4 & $0$ & $0.400$ & $0.143$ & -$5$ \\ 
5 & $1$ & $0.500$ & $0.143$ & $0$ \\ 
6 & $2$ & $0.600$ & $0.143$ & $5$ \\ 
7 & $3$ & $0.700$ & $0.143$ & $5$ \\ 
\hline \\[-1.8ex] 
\end{tabular}  
    \end{subfigure}%
    \begin{subfigure}{0.5\textwidth}
       \includegraphics[width=0.9\textwidth,  valign=m]{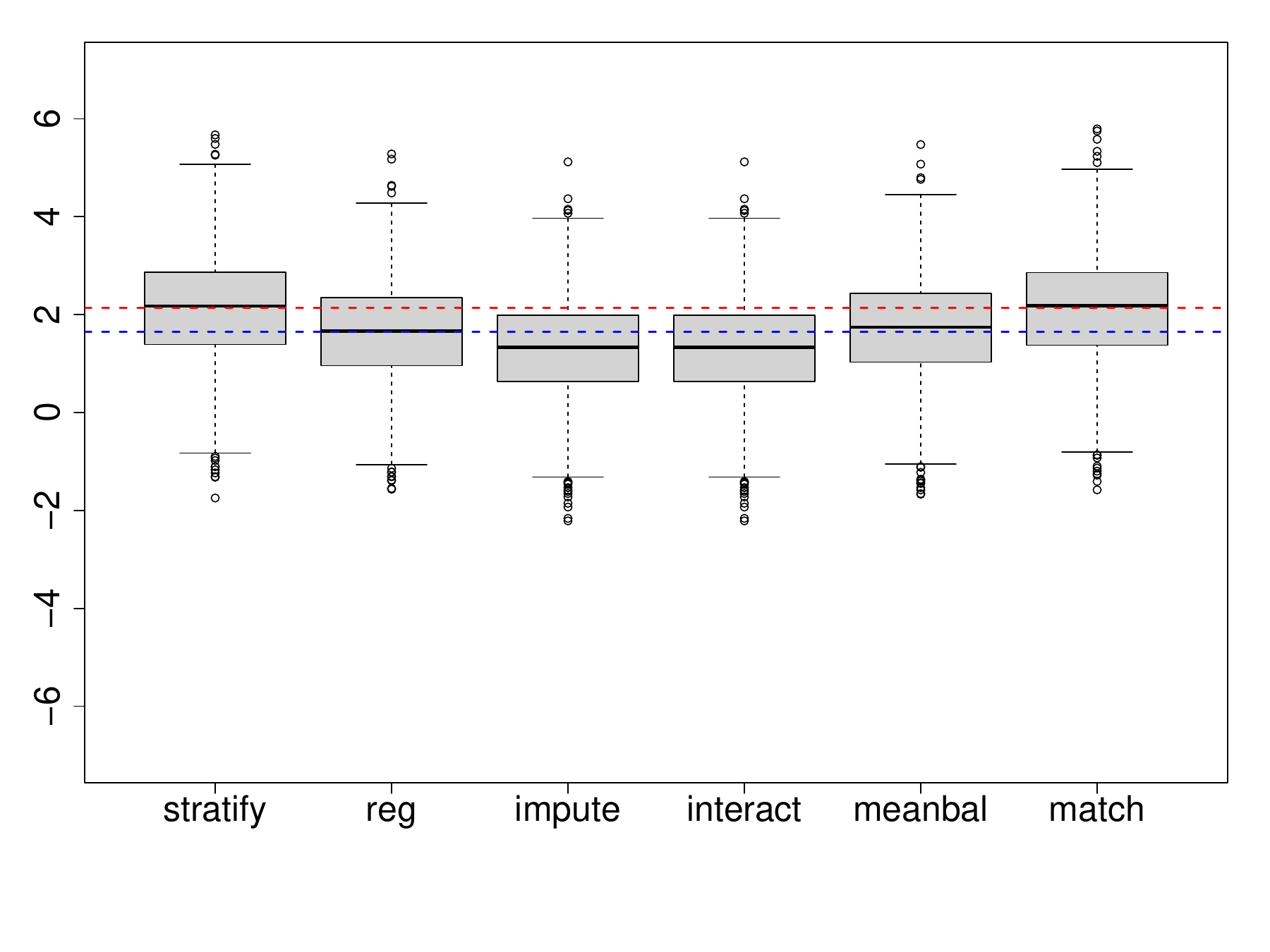} 
    \end{subfigure}

      \caption{Estimates over 500 iterations of size $n=1000$. The dashed red line indicates the true ATE. The dashed blue and green lines show the (expected) regression coefficients reconstructed using  the \cite{angrist_empirical_1999} weights (blue), or our more general weights (green).}
    \label{fig:discrete-sim-results}
\end{figure}

The first three DGPs involve a binary treatment variable $D$, a discrete covariate $X$ in the range $[-3, 3]$, and an outcome $Y$ which depends on $D$ and $X$. For each simulation setting, the tables in Figure \ref{fig:discrete-sim-results} show the possible values of $X$, the corresponding probability of treatment, probability that $X=x$, and the average treatment effect $\tau_x$ for the subgroup where $X=x$.  In all simulations, noise is added to the outcome to achieve an $R^2$ of 0.33 between the systematic (noiseless) portion of $Y$ and the final $Y$ with noise.

As expected, when there is a high level of heterogeneity in treatment effect and in probability of treatment, the regression adjustment estimate will have a substantial amount of bias. Figure \ref{fig:discrete-sim-results} shows the effect estimates when there is heterogeneity in both treatment effect and probability of treatment between subgroups. In the first two settings, the potential outcomes are linear in $X$. Here we see that the OLS estimate (\texttt{reg}) is heavily biased, and would lead investigators to conclude there is a statistically significant positive effect, though the true ATE is zero. Notably, slightly more extreme simulations would even make it possible for OLS to produce the incorrect sign for the treatment effect estimate. Meanwhile, regression imputation (\texttt{impute}), the Lin estimator (\texttt{interact}), mean balancing (\texttt{meanbal}), and matching (\texttt{match}) all successfully address this concern. We recommend the use of regression imputation or equivalently the Lin/interacted adjustment as a simple way to improve on conventional OLS estimation. 

Naturally, in the case where the potential outcomes are not linear in the treatment and covariates, the estimates from the interacted model, imputation, and mean balancing are all biased for the ATE. This is expected, as the assumptions of both single and separate linearity are violated. Addressing such non-linearity requires non-linear estimators, such as matching.  We also note that the 
variance-weighted estimate (using Expression~\ref{varweight-equation}) does not always reproduce the actual OLS estimate, as in the first simulation setting, where $P(D=1|X)$ is not linear in $X$. This is easily avoided by fully saturating the model in $X$, although such a manuever would not work for the continuous cases below.  

The more general (but infeasible) weighting representation (Expression~\ref{trueweights}) reproduces the OLS estimate exactly regardless of the functional form of $P(D_i|X_i)$. 

\subsubsection*{Standard errors}

While these results show the expected variability in estimates under resampling from the given DGP, for an investigator working with one observed dataset, some form of \textit{estimated} standard error is vitally important to inference.  Table~\ref{tab.ses.dgp1.discrete} reports the average analytically estimated standard errors for DGP1 above, still with discrete $X$. Results are similar in other settings. 

For stratification, we take a weighted sum of strata-specific Neyman variances. Here and below we write expressions in the plug-in/analog sample estimator form.
\begin{align}
    \widehat{\V}(\hat{\tau}_{strat}) = \sum_{x \in X} \hat{p}(X=x)^2 \left( \frac{\widehat{\V}(Y_i | D=1, X=x)}{n_1} + \frac{\widehat{\V}(Y_i | D=0, X=x)}{n_0} \right)
\end{align}

For the simple and interacted regression, we calculate the HC2 standard error of the coefficient on the treatment variable. For regression imputation, we calculate the standard error again in the Neyman style, 

\begin{align}
    \widehat{\V}\left(\frac{1}{n} \sum_{i} (\hat{\mu}_1(X_i) - \hat{\mu}_0(X_i))\right) &= \widehat{\V}\left(\frac{1}{n} \sum_{i} (\gamma X_i - \alpha X_i)\right) \\
    &= \widehat{\V}(\gamma\bar{X} - \alpha\bar{X}) \\
    &= \bar{X}^T \widehat{\Sigma}_1 \bar{X} + \bar{X} \widehat{\Sigma}_0 \bar{X}
\end{align}

\noindent where $\hat{\Sigma}_1$ and $\hat{\Sigma}_0$ are the estimated variance-covariance matrices for the treatment model and the control model, respectively. 

For mean balancing we show two types of analytical standard errors. First, we use the (HC2) standard error from the weighted regression of just $Y$ on $D$. Second, \texttt{meanbal-adj} uses the HC2 standard errors from a regression of $Y$ on $D$ and $X$, again with the estimated weights. Both methods produce identical point estimates (when perfect mean balance is achieved by the weights), but the analytical standard errors of the \texttt{meanbal-adj} approach benefit from partialing out the $X$. This is akin to how conventional OLS standard errors, under a fixed design, partial $X$ out of $Y$ so that the estimates are built on the conditional/residual variance of $Y$ rather than the total variance. For matching we use the Abadie-Imbens standard error \citep{abadie_large_2006}.

\begin{table}[!htbp] \centering 
 
\begin{tabular}{@{\extracolsep{5pt}} ccccc} 
\\[-1.8ex]\hline 
\hline \\[-1.8ex] 
 & bias & rmse & avg analytical SE & empirical SE \\ 
\hline \\[-1.8ex] 
stratify & -$0.021$ & $1.431$ & $1.478$ & $1.432$ \\ 
reg & $2.021$ & $2.376$ & $1.264$ & $1.250$ \\ 
impute & $0.005$ & $1.315$ & $1.321$ & $1.316$ \\ 
interact & $0.005$ & $1.315$ & $1.321$ & $1.316$ \\ 
meanbal & -$0.001$ & $1.328$ & $1.812$ & $1.329$ \\ 
meanbal-adj & -$0.002$ & $1.328$ & $1.379$ & $1.329$ \\ 
match & -$0.005$ & $1.434$ & $1.514$ & $1.436$ \\ 
\hline \\[-1.8ex] 
\end{tabular} 
\caption{Simulation results under DGP1. The bias and rmse columns repeat information shown graphically in Figure~\ref{fig:discrete-sim-results}. The empirical SE gives the actual standard deviation of the estimates across resamples. The average analytical SE is the average of the standard errors that would have been computed analytically in each iteration, using the estimators described in the text.} \label{tab.ses.dgp1.discrete} 
\end{table} 

We find, first, that the empirical standard errors are extremely similar across the methods relying on single or double linearity (\texttt{reg}, \texttt{impute}, \texttt{interact}, \texttt{meanbal}, \texttt{meanbal-adj}). The approaches not relying on single or separate linearity (\texttt{stratify} and \texttt{match}) show somewhat larger standard errors, as expected given their greater flexibility. Second, each method's average analytical standard is within 5\% of the empirical standard error, with the exception of \texttt{meanbal}, with an average analytical standard error almost 40\% larger than the empirical value. This is repaired, however, by using \texttt{meanbal-adj}.

Finally, some investigators may be concerned with efficiency in the sense of sampling variability in the estimate and its consequences for inference. This can be understood as a bias-variance tradeoff. The comparison between OLS (\texttt{reg}) and the Lin/interaction approach (\texttt{interact}) offers a simple starting point, since the later will add one additional parameter to the regression for every covariate dimension. Because the number of observations is large relative to the number of covariates, this has very little impact on the estimate's variability across resamples. For example in DGP 1, Table~\ref{tab.ses.dgp1.discrete} shows that the empirical SE is only about 5\% larger for \texttt{interact} than for \texttt{reg}. The behavior of \texttt{impute} is of course identical. The empirical SE from \texttt{meanbal} and \texttt{meanbal\-adj} are similarly only 6\% larger than from \texttt{reg}. If an investigator is primarily concerned with root mean square error (RMSE) of the estimates around the true value, RMSE values fall by nearly half for each of these methods relative to \texttt{reg}.  That is, the small loss of efficiency in these settings (increasing variance) is more than made up for by reductions in bias as they factor into the RMSE.

It is important to recognize, however, the  favorable nature of our simulation setting in this regard.  If the number of covariates was large enough relative to the sample size, if treatment probability varied little by stratum of $X$, and/or if efficiency  was a greater concern than bias or RMSE, then investigators might have cause to prefer OLS and adopting its weighted-ATE as the target estimand for their inferential purposes. 

\subsection{Continuous covariate}
\begin{figure}[t!]
\begin{subfigure}[t]{0.3\textwidth}
    \centering
    \captionsetup{width=.8\linewidth}
    \includegraphics[width=\linewidth]{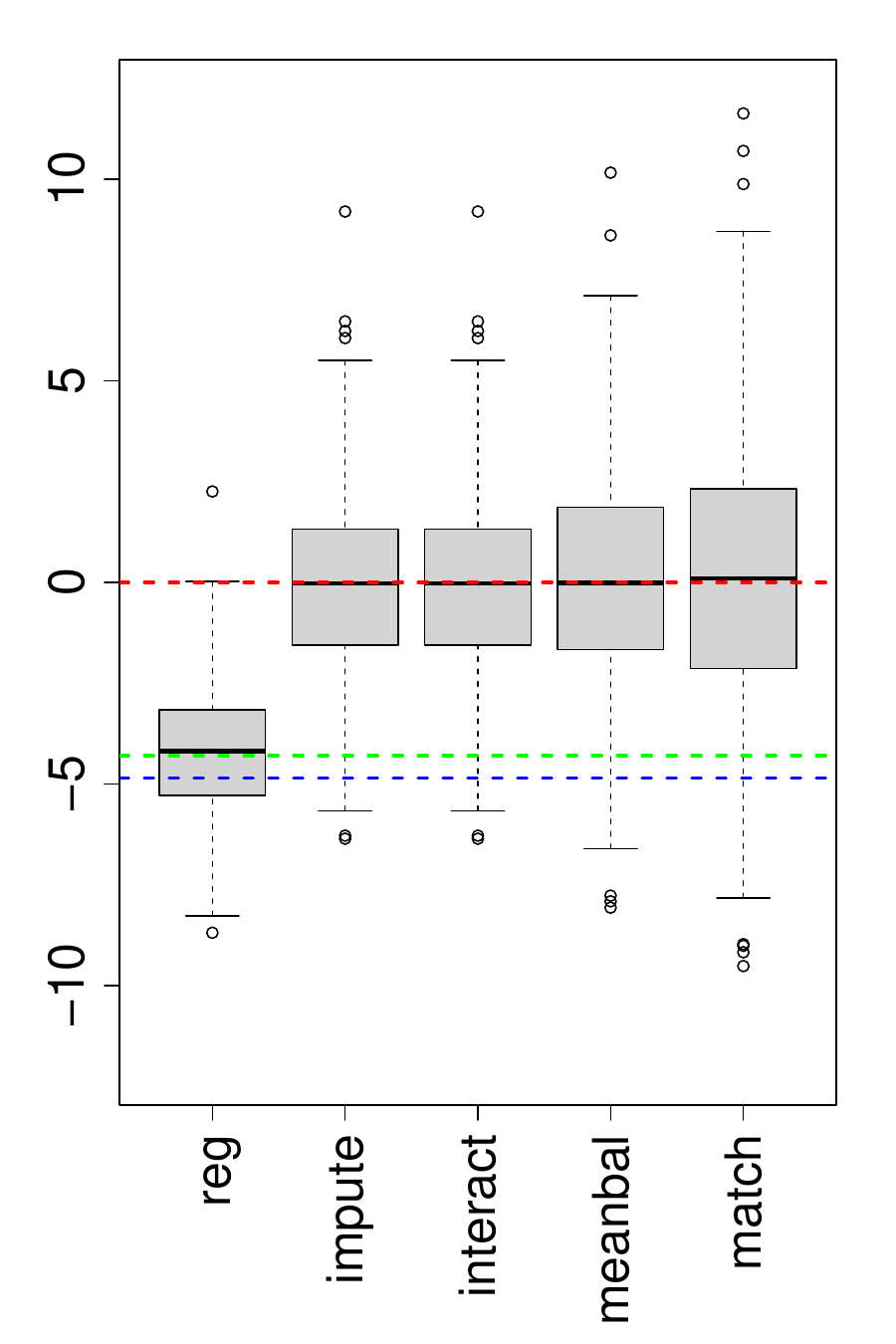}
    \caption{X is randomly sampled from Unif[-3,3], $P(D \vert X) = \frac{1}{1 + e^{-2-x}}$, and $Y(1) = Y(0) + 3X$}
    \label{fig:cont_threshold}
\end{subfigure}
\hfill
\begin{subfigure}[t]{0.3\textwidth}
    \centering
    \captionsetup{width=.8\linewidth}
    \includegraphics[width=\linewidth]{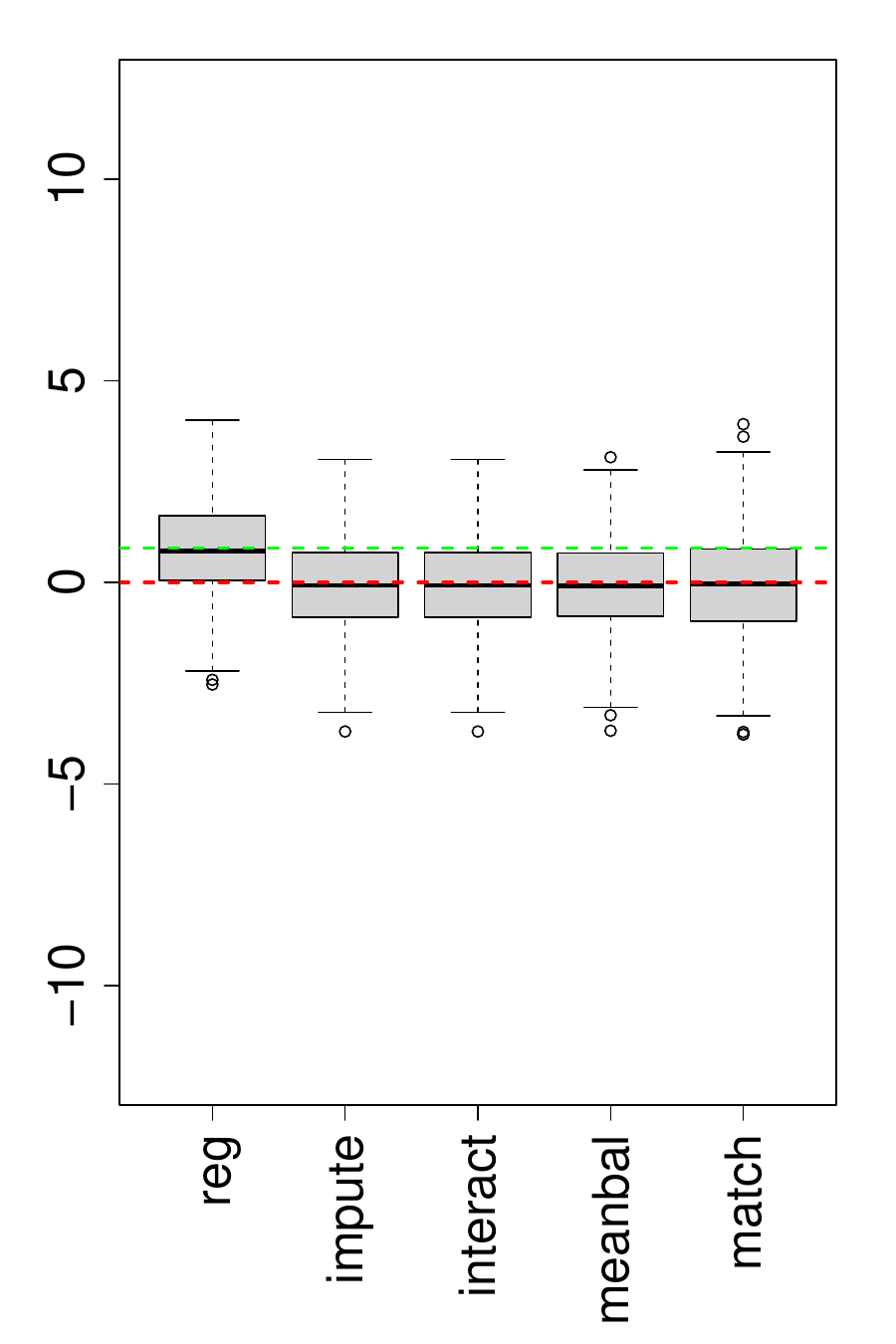}
    \caption{X is randomly sampled from Unif[-3,3], $P(D \vert X) = 0.1X + 0.4$, and $Y(1) = Y(0) + 3X$}
    \label{fig:cont_linear}
\end{subfigure}
\hfill
\begin{subfigure}[t]{0.3\textwidth}
    \centering
    \captionsetup{width=\linewidth}
    \includegraphics[width=\linewidth]{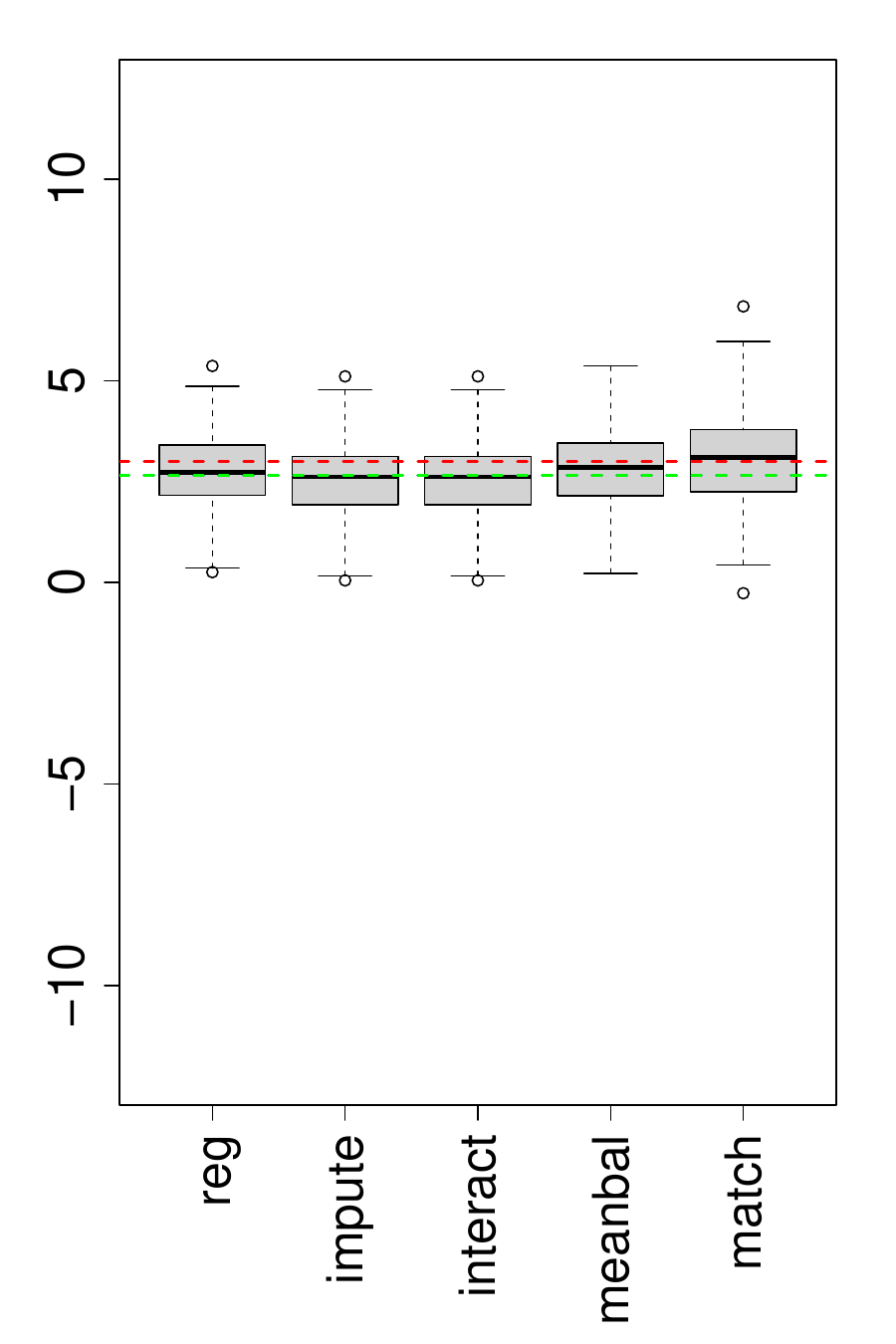}
    \caption{X is randomly sampled from Unif[-3,3], $P(D \vert X) = 0.1X + 0.4$, and $Y(1) = Y(0) + X^2$.}
    \label{fig:cont_nonlinear}
\end{subfigure}
    
    \caption{Effect estimates for 200 samples of size n=1000 under data generating processes where $X$ is continuous and different estimation strategies. The dashed red line is set at the true ATE. The dashed blue line shows the average (over population) value for the variance-weighted formulation of the weights per \cite{angrist_empirical_1999}. The dashed green line shows the (expected) regression coefficients reconstructed using our more general weights.}
    \label{fig:main-results}
\end{figure}

While we have considered discrete $X$ thus far, we do so for the sake of intuition regarding strata, but the lessons apply to settings with continuous $X$ as well. In the simulations shown in Figures \ref{fig:cont_threshold} and \ref{fig:cont_linear}, $X$ is randomly sampled from uniformly from $[-3, 3]$, and $Y(1) = Y(0) + 3X$. However, $P(D|X)$ takes on a different form in each setting. For the first continuous specification, $P(D \vert X)$ takes a logistic form. In the second, $P(D \vert X)$ increase linearly with $X$.  For the third specification, $P(D \vert X)$ increases linearly in $X$, but $Y(1)$ is nonlinear in $X$. 

In each of these settings, we see that the OLS estimate is biased for the true ATE. 
As before, the interaction, imputation, and mean balancing estimators perform well, except in Figure~\ref{fig:cont_nonlinear} where even separate linearity fails.

\subsection{Covariate adjustment in experiments, block randomization, and baseline-free average marginal effects}\label{blockrandomization}

These lessons may at first seem to be directed to researchers working with observational data such that conditioning on $X$ is a requirement. However, they also apply in settings where conditioning on $X$ is used to improve precision/ minimize finite sample differences from the expected result. Our advice essentially echoes that of \cite{lin_agnostic_2013} and a broad subsequent literature calling for a model that interacts treatment with the (centered) covariates. As noted above, this is identical to analogous imputation/g-computation/T-learner approaches when using OLS for the underlying models. 

Consider first the cases where investigators have a randomized experiment, but adjust for covariates in the analysis to improve precision. When $D$ is fully randomized, the probability of treatment across values of $X$ will typically not vary greatly, except by chance. This implies that the bias due to regression's weighting behavior will typically be small. Nevertheless, any discrepancy between the ATE and coefficient on account of these weights can be avoided entirely. 

Second, block randomization is a powerful tool often employed by experimentalists to reduce finite sample deviations from expectation. For the estimates to reflect the reduced variance this affords, it is standard to regress the outcome ($Y$) on the treatment ($D$) and block indicators ($B$) as fixed effect covariates. The weights implied by regression of $Y$ on $D$ and block indicators $B$ would be given by

\begin{align}
    \hat{\tau}_{BFE} = \frac{\sum_{b=1}^{P} \hat{\tau}_b P(D_i = 1 | B_i = b) (1 - P(D_i=1 | B_i = b)) P(B_i = b)}{\sum_{b=1}^{P} P(D_i = 1 | B_i = b) (1 - P(D_i=1 | B_i = b)) P(B_i = b)}
\end{align}

\noindent where it is innocuous to assume that $P(D=1|B)$ is linear in the block indicators, as this  regression would be fully saturated in $B$. 
If all blocks have the same probability of treatment, the weights are equal, so the regression coefficient will represent the difference in means per block, averaged over the size of each block, i.e. the ATE. This will often be the case. However, if the treatment probability varies by block, either by design or otherwise, then the resulting coefficient estimate would not in general be the ATE. 
Rather, regression will put higher weight on blocks with probabilities of treatment nearer to 50\%. 

As above, this is simply a consequence of misspecification generated by heterogeneous treatment effect estimates by block.  Including interactions between the treatment and the block fixed effects would address this, under the separate linearity assumption,
\begin{align}
    \E[Y_i(1) | B_i] &= \alpha_0 + \alpha_1 \mathds{1}
\{B_i = 1\} + .. + \alpha_b \mathds{1} \{B_i = P\} \\
    \E[Y_i(0) | B_i] &= \gamma_0 + \gamma_1 \mathds{1}\{B_i = 1\} + ... + \gamma_b \mathds{1} \{B_i = P\}
\end{align}

One complication when applying the interacted approach here is that care must be taken regarding the interpretation due to the interaction. In typical usage, one block indicator will be omitted to avoid co-linearity with the intercept. If no centering/de-meaning is done on the block indicators, the coefficient estimate would represent the estimated effect (difference in means) in whichever block had its indicator omitted. The solution of centering covariates \citep{lin_agnostic_2013} as utilized above is now complicated by this omission. It is possible to omit the intercept, rather than the indicator for one block, and utilize the centering. However, a more general solution is to avoid any consideration of centering and omitting one level/the intercept, and works when dealing with one or multiple categorical variables. This is to simply compute the marginal effect estimate for each observation, in whatever block it is in, $\frac{\partial Y}{\partial D}\big|_{B=b}$. Averaging these across observations (giving equal weight to each observations) produces the average marginal effect (AME),

\begin{align}
\hat{\tau}_{AME} &= \frac{1}{n} \sum_{i=1}^n \widehat{\frac{\partial Y}{\partial D}}\bigg|_{B=b}
\end{align}
\noindent where $\widehat{\frac{\partial Y}{\partial D}}\big|_{B=b}$ is the estimated marginal effect in block $b$ where this individual unit is found. For example, in the regression
\begin{align}
Y_i = \beta_0 + \tau D &+ \beta_1 \mathds{1}
\{B_i = 1\} + \beta_2 \mathds{1} \{B_i = 2\} + ... +  \beta_P \mathds{1}
 \{B_i = P\} \\
 &+ \alpha_1 \mathds{1}
\{B_i = 1\}D + \alpha_2 \mathds{1} \{B_i = 2\}D + ... +  \alpha_P \mathds{1}
 \{B_i = P\}D + \epsilon_i 
 \end{align}

\noindent the estimated marginal effect in block $B=b$ is $\hat{\tau} + \alpha_p$, and so the average marginal effect of interest over the whole sample is simply 
\begin{align}
    \hat{\tau}_{AME} = \hat{\tau} + \frac{1}{n} \sum_{i=1}^n \sum_{b=1}^P \left(\hat{\alpha}_p \mathds{1}\{B_i=b\}\right)
\end{align}

Using this approach to interpret the fitted regression with interactions always produces an estimate with the desired interpretation of ``the estimated marginal effect, which can differ by block, averaged over blocks according to how many observations fall in each.'' In the case of block indicators or other categorical variables, this approach avoids errors in relation to choices about what levels to omit in the regression, whether to omit the intercept, or having to center these variables. The variance for this estimator is
\begin{align}
    \V(\hat{\tau}_{AME}) = \V(\hat{\tau}) + \sum_{p=1}^P P(B_i = p)^2 \V(\hat{\alpha}_p) &+ 2 \sum_{k < j} P(B_i = k) P(B_i = j) \Cov(\hat{\alpha}_k, \hat{\alpha}_j) \\ &+ 2 \sum_{p=1}^P P(B_i = p) \Cov(\hat{\tau}, \hat{\alpha}_p)
\end{align}
For comparison, we also consider two other approaches. One practice is to weight the blocked fixed effects regression by the (stabilized) inverse probability of treatment assignment for each block, 
\begin{align}
    w_{i}^{\text{IPW}} = \frac{P(D_i=1)}{P(D_i=1|B=b_i)} D_i + \frac{P(D_i=0)}{P(D_i=0|B=b_i)} (1 - D_i) 
\end{align}
These weights are constructed so that within each block, the treated and control units make up equal weighted proportions. They therefore neutralize any differences in the $P(D=1|B)$ across blocks. If we perform a weighted regression of $Y$ on $D$ and the block dummy variables using these weights, the coefficient estimate for $D$ will be unbiased for the ATE.\footnote{A weighted difference in means (rather than regression including $B$) with these weights would produce the same point estimate, but the improvement in efficiency obtained under the block randomization design will not be fully reflected in the estimated standard error. This occurs because the block indicators will be orthogonal to treatment (once weighted) and so do not affect the coefficient estimate on treatment, but the omission of $B$ prevents the model from reducing the residuals that enter the standard error.}

Finally we also the stratification estimator (Expression~\ref{avedim}) but with blocks as strata, 
\begin{align}
    \hat{\tau}_{\text{blockDIM}} = \sum_{b = 1}^{|B|} \hat{P}(B = b) (\bar{Y}_{D=1, B=b} - \bar{Y}_{D=0, B=b}).
\end{align}

\begin{figure}[htb!]
    \centering\includegraphics[width=.7\textwidth]{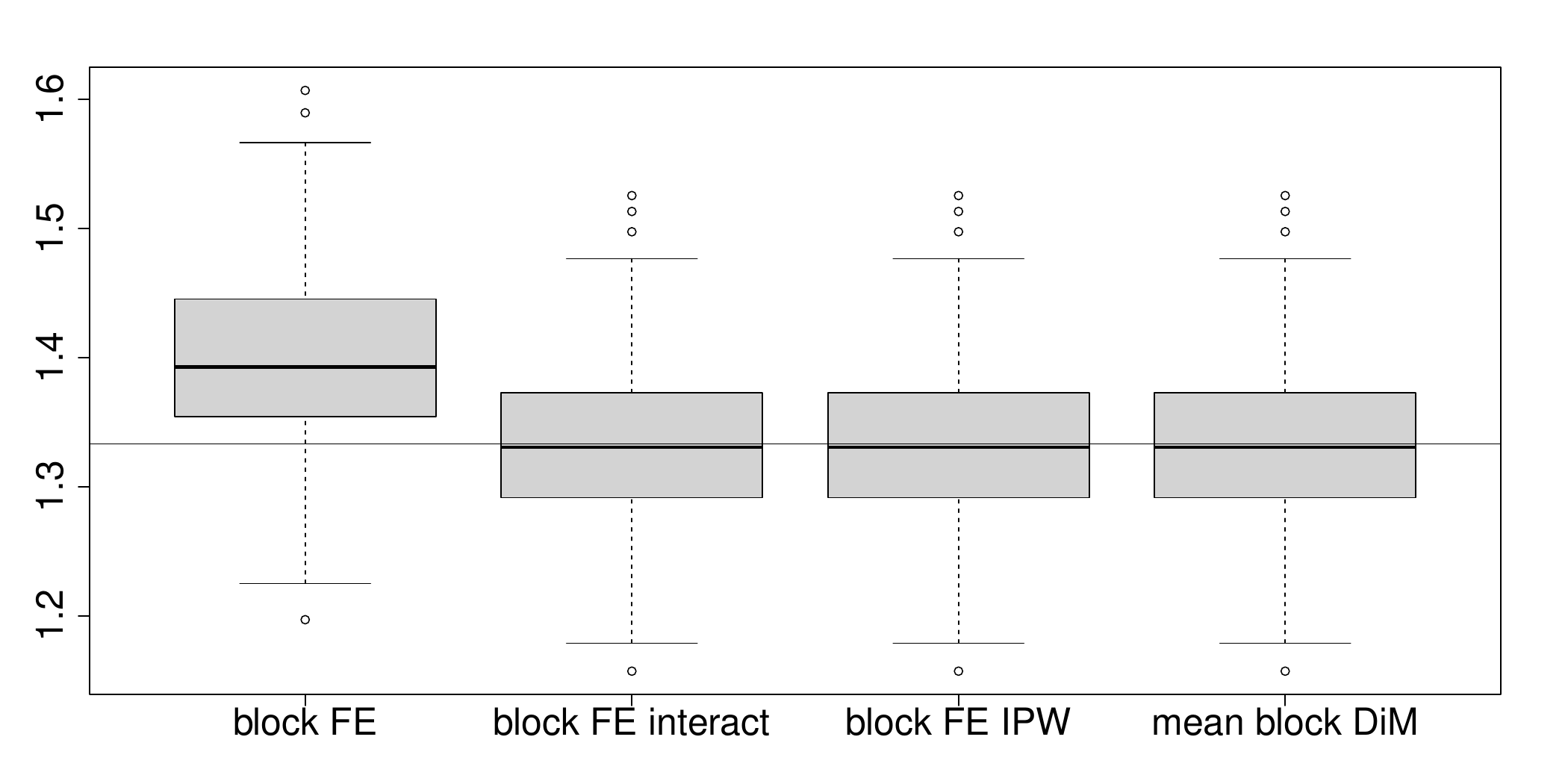}
    \caption{Effect estimates for 500 iterations of size $n=1200$ each. $D$ is assigned by (complete) randomization within each block with probability $P(D|B)$, and $Y = 2 + D^*\tau$. For blocks 1-6, $P(D|B)$ = (0.25, 0.25, 0.5, 0.25, 0.25, 0.5) respectively, and $\tau = 4^*P(D|B)$. The black line indicates the true ATE.}
    \label{fig:blockrand-results}
\end{figure}

Figure \ref{fig:blockrand-results} compares estimates from the ``plain'' block fixed effects (\texttt{block FE}) regression to the interacted regression with the average marginal effect interpretation (\texttt{block FE interact}), the IPW-weighted regression (\texttt{block FE IPW}), and the block DIM average (\texttt{mean block DIM}) in a simulation setting where treatment probability varies by block and there is heterogeneity in treatment effect. As expected, the block fixed effects OLS model suffers from the weighting problem. All three alternative approaches produce unbiased and identical estimates. 

We recommend the use of one of these alternative estimators to improve upon estimation of the ATE under block randomization.\footnote{These results are  consistent with those demonstrated in \href{https://declaredesign.org/blog/posts/biased-fixed-effects.html}{https://declaredesign.org/blog/posts/biased-fixed-effects.html}, which use the DeclareDesign simulation approach \citep{blair2019declaring} and conclude that the mean blockwise DiM/ stratification approach is suitable.}

\section{Conclusion}

Given the wide use of OLS regression of $Y$ on $D$ and $X$ in practice, and the potential gap between the coefficient and target quantities like the ATE, prior work has provided several forms of guidance to practitioners regarding this weighting problem. \cite{humphreys_bounds_2009} notes the conditions under which the OLS result will be nearer to ATT or ATC, and observes monotonicity conditions under which the coefficient estimate will fall between these two endpoints. \cite{aronow_does_2016} suggest characterizing the ``effective sample'' for which the regression coefficient would be an unbiased estimate of the ATE by applying the variance weights to the covariate distribution. Other papers have provided diagnostics for quantifying the severity of this bias. \cite{chattopadhyay2023implied} characterize the implied unit-level weights of regression adjustment (as we do as an intermediate result), and propose diagnostics that use these weights to analyze covariate balance, extrapolation outside the support, variance of the estimator, and effective sample size. These ideas relate closely to those of \cite{aronow_does_2016} in that they compare the covariate distribution targeted by the estimator to that of the target population. \cite{sloczynski_interpreting_2022} derives a diagnostic related to variation in treatment probability, 
\begin{equation}
    \delta =  \frac{\rho^2 \V[P(D|X) | D=1] - (1 - \rho)^2 \V[P(D|X)|D=0]}{\rho \V[P(D|X) | D=1] + (1 - \rho) \V[P(D|X) | D=0]}
\end{equation}
where $\rho$ is the unconditional probability of treatment. The bias due to heterogeneity will be
$\delta (\tau_{ATC} - \tau_{ATT})$ where $\tau_{ATC}$ and $\tau_{ATT}$ are the average treatment effect among treated and control, respectively.  Thus $\delta$ captures one aspect of the potential for bias---variation in treatment probability---but reasoning about the severity of bias requires knowledge of the heterogeneity in treatment effects, which is unknown to the investigator.

However, investigators could instead consider different modeling choices that avoid this ``weighting problem'' altogether. We have sought to clarify the theoretical considerations required to do so, first by understanding the weighting problem as a symptom of misspecification when the treatment effect (and probability of treatment) vary with $X$. We provide a more general expression for the ``weights of regression'' by algebraically manipulating a representation of the OLS coefficient. The key fact is that the apparent ``weights of regression'' simply account for the way misspecification alters the coefficient away from the ATE. Second, viewed this way, a natural proposal for research practice is to rely instead on specifications that will not necessarily be violated by effect heterogeneity.  To focus on minimal changes in practice, the conventional \textit{single linearity} assumption (that $Y$ is linear in $D$ and $X$) can be relaxed at least to the \textit{separate linearity} assumptions (each potential outcome is separately linear in $X$). This recapitulates equivalent assumptions employed in analyses by \cite{hahn2023properties, sloczynski_interpreting_2022, kline2011} and \cite{imbens_recent_2009}. It also clarifies which estimators are suitable to avoid these weights, and whose properties we can consider by comparison. Fortunately, a number of existing, straightforward estimation approaches produce unbiased ATE estimates under this assumption. First, regression imputation (including g-estimation and the T-learner) with OLS as well as including the interaction of $X$ and $D$ (as in \citealp{lin_agnostic_2013}) are all suitable, and identical to each other in this setting. This longstanding approach dates back to at least \cite{peters1941method} and \cite{belson1956technique}, as noted by \cite{Chattopadhyay2024Causation} who label it as ``multi-regression''. In addition, ``mean balancing'' estimators applied to the ATE---those that choose weights to achieve the same mean of $X$ for the treated and control group as in the full sample---are also unbiased under separate linearity. These are not identical to the above as they avoid fitting any model, and our simulations suggest this property can partially mitigate bias when even the separate linearity assumption fails. These approaches can be useful not only in observational studies (so far as investigators can claim conditional ignorability), but also when using covariate adjustment after randomization, including the special case of analyzing block randomized experiments.  

Avoiding these weights through these alternatives may be desirable in many cases, but is not without cost of limitation. The natural cost to first consider (relative to regression under single linearity) is the potential loss in efficiency. The interact/imputation/g-computation/T-learner approach adds an additional nuisance parameter per covariate. The circumstances investigated here---with just one covariate, a high degree of treatment effect heterogeneity, and large variation in treatment probability---clearly favor these approaches, where they eliminate bias and reduce RMSE by roughly half, while increasing the standard error by only about 5\% in our settings. However, there may be settings in which the improved bias does not so assuredly dominate the efficiency cost. For example, where there are many covariates relative to the sample size,  little difference in the probability of treatment by stratum, or little theoretical reason to expect treatment effect heterogeneity, investigators may have cause to prefer OLS if efficiency concerns are of paramount interest. In those cases, diagnostics or interpretational aids that gauge the severity of the misspecification bias (weights) may be useful, depending on how investigators prefer to tradeoff bias and efficiency in their inferential or decision-making goals.

Finally, when the linearity in potential outcomes assumption is not satisfied, none of these methods can guarantee unbiasedness. Our work here regards only the relaxation from single linearity to separate linearity. Investigators not satisfied with the linear approximation to treatment effects in this way could reasonably consider more flexible approaches---though doing so may incur additional uncertainty costs as illustrated in Table~\ref{tab.ses.dgp1.discrete}. Nevertheless, we agree with assessments such as \cite{keele_adjusting_2010}  and \cite{aronow_does_2016} that current practice largely remains reliant on linear approximations to adjust for covariates while hoping to interpret the result as a meaningful causal effect. Our analysis suggests that, especially where there are enough observations relative to covariates to support imputation/g-computation/T-learner, interactive regression, or mean balancing, this may be preferable to suffering regression's ``weighting problem'', as it avoids this form of bias under slightly weaker assumptions and requires only a small change to current practice.

\bibliographystyle{plainnat}
\bibliography{references}


\newpage

\appendix
\setcounter{page}{1}  

\part{Appendix - Demystifying and avoiding the OLS ``weighting problem''} 
\parttoc 

\clearpage

\section{Appendix}

\subsection{Derivation of more general weights}
\label{derive-realweights}
The above weights rely on the assumption that $\E[D_i | X_i]$ is linear. We can derive more general weights that do not involve this assumption by using a more general representation of $D^{\perp X}$:

\begin{equation}
    \hat{\tau}_{reg} = \frac{\widehat{\Cov}(Y_i, D_i^{\perp X})}{\widehat{\V}(D_i^{\perp X})} 
\end{equation}


\begin{equation}
    D^{\perp X} = D - \X (\X^T \X)^{-1} \X^T D = D - \X \theta
\end{equation}

where $\X = \begin{bmatrix}
\vec{1} & \vec{X} \end{bmatrix}$. We first derive the unit-wise weights, which are somewhat trivial.

\begin{align}
    \widehat{\Cov}(Y_i, D_i^{\perp X}) &= 
    \hat{\E}[Y_i D_i^{\perp X}] - \hat{\E}[Y_i] \hat{\E}[D_i^{\perp X}] \\
    &= \hat{\E}[Y_i D_i^{\perp X}] \\
    &= \sum_i Y_i (D_i - \X_i \theta) \\
    &= \sum_i Y_i (D_i - \hat{d}(X_i)) \\
\end{align}
where $\hat{d}(X_i) = \X_i \beta$ where $\X_i = \begin{bmatrix} 1 & X_i \end{bmatrix}$ and $X_i = x$.

\begin{align}
    \widehat{\V}(D_i^{\perp X}) &= \hat{\E}[D_i^{{\perp X}^2}] - \hat{\E}[D_i^{\perp X}]^2 \\
    &= \hat{\E}[D_i^{\perp X^2}] \\
    &= \sum_i (D_i - \X_i \theta)^2 \\
    &= \sum_i (D_i - \hat{d}(X_i))^2 \\
\end{align}
So, the unit-wise weighted representation is as follows:
\begin{equation}
    \hat{\tau}_{reg} = \frac{\widehat{\Cov}(Y_i, D_i^{\perp X})}{\widehat{\V}(D_i^{\perp X})} = \frac{\sum_i Y_i (D_i - \hat{d}(X_i))}{\sum_i (D_i - \hat{d}(X_i))^2}
\end{equation}
Let $\pi_x = P(D_i=1 | X_i = x)$ and $\hat{d}(X_i) = \X_i \theta$ where $\X_i = \begin{bmatrix} 1 & X_i \end{bmatrix}$ and $X_i = x$. We can try to manipulate the above representation to get the strata-wise weights:
\begin{align}
\widehat{\Cov}(Y_i, D_i^{\perp X})
    &= \sum_x P(X=x) \hat{\E}[Y_i (D_i - \hat{d}(X_i)) | X_i] \\
    &= \sum_x P(X=x) \hat{\E}[Y_i D_i | X_i] - \hat{d}(X_i) \hat{\E}[Y_i | X_i] \\
    &= \sum_x P(X=x) (P(D_i=1 | X_i) \hat{\E}[Y_i | D_i=1, X_i] - \hat{d}(X_i) \hat{\E}[Y_i | X_i]) \\
    &= \sum_x P(X=x) (\pi_x \hat{\E}[Y_i | D_i=1, X_i] 
    - \hat{d}(X_i) (\pi_x \hat{\E}[Y_i|D_i=1, X_i] \\
    &+ (1 - \pi_x) \hat{\E}[Y_i | D_i=0, X_i)) \nonumber \\
    &= \sum_x P(X=x) (\pi_x (1 - \hat{d}(X_i)) \hat{\E}[Y_i | D_i=1, X_i] 
    - \hat{d}(X_i) (1-\pi_x) \hat{\E}[Y | D_i=0, X_i])
\end{align}
This can be rewritten in a couple different ways, where $\tau_x$ is the ATE given $X=x$:
\begin{align}
    \widehat{\Cov}(Y_i, D_i^{\perp X}) &= \sum_x P(X=x) (\hat{\E}[Y_i | D_i=1, X_i] (\pi_x(1 - \hat{d}(X_i))) - \hat{\E}[Y | D_i=0, X_i](\hat{d}(X_i) (1 - \pi_x)) \\
    &= \sum_x P(X=x) (\pi_x \hat{\E}[Y_i | D_i=1, X_i] - \hat{d}(X_i) \hat{\E}[Y_i|D_i=0, X_i] - \hat{d}(X_i) \pi_x \tau_x \\
    &= \sum_x P(X=x) \pi_x (1 - \hat{d}(X_i)) (\hat{\E}[Y_i | D_i=1, X_i] - \frac{\hat{d}(X_i)}{1 - \hat{d}(X_i)} \frac{1 - \pi_i}{\pi_i} \hat{\E}[Y_i | D_i=0, X_i])
\end{align}

Next, to get the full expression for $\hat{\tau}_{reg}$, we calculate $\V(D_i^{\perp X})$.
\begin{align}
    \widehat{\V}(D_i^{\perp X}) &= \sum_x P(X=x) \hat{\E}[D_i^{{\perp X}^2} | X] \\
    &= \sum_x P(X=x) \hat{\E}[D_i | X_i] - 2 \X_i \beta \hat{\E}[D_i | X_i] + (\X_i \beta)^2 \\
    &= \sum_x P(X=x) (\pi_x - 2 \hat{d}(X_i) \pi_x + \hat{d}(X_i)^2) \\
    &= \sum_x P(X=x) (\pi_x(1 - \hat{d}(X_i)) + \hat{d}(X_i)(\hat{d}(X_i) - \pi_x))
\end{align}

So we end up with the following expression for $\hat{\tau}_{reg}$.

\begin{equation}
\label{ref:appendix-trueweights}
    \hat{\tau}_{reg} = \frac{\sum_x P(X=x) (\hat{\E}[Y_i | D_i=1, X_i] (\pi_x(1 - \hat{d}(X_i))) - \hat{\E}[Y | D_i=0, X_i](\hat{d}(X_i) (1 - \pi_x))}{\sum_x P(X=x) (\pi_x(1 - \hat{d}(X_i)) + \hat{d}(X_i)(\hat{d}(X_i) - \pi_x))}
\end{equation}

Suppose we let $\hat{d}(X_i) = \pi_x + a_x$, where $a_x$ is the difference between the true probability of treatment given $X=x$ and the probability of treatment estimated by a linear model. We can then write $\widehat{\Cov}(Y_i, D_i^{\perp X})$ in terms of $a_x$:
\begin{align}
    \widehat{\Cov}(Y_i, D_i^{\perp X}) &= \sum_x P(X=x) \pi_x \hat{\E}[Y_i | D_i=1, X_i] - \hat{d}(X_i) \hat{\E}[Y_i | D_i=0, X_i] - \hat{d}(X_i) \pi_x \tau_x \\
    &= \sum_x P(X=x) \pi_x \hat{\E}[Y_i | D_i=1, X_i] - (\pi_x + a_x) \hat{\E}[Y_i | D_i=0, X_i] - (\pi_x + a_x) \pi_x \tau_x \\
    &= \sum_x P(X=x) \pi_x \hat{\E}[Y_i | D_i=1, X_i] - \pi_x \hat{\E}[Y_i | D_i=0, X_i] - a_x \hat{\E}[Y_i | D_i=0, X_i] \\
    &- \pi_x^2 \tau_x - a_x\pi_x \tau_x \nonumber \\
    &= \sum_x P(X=x) \pi_x (1 - \pi_x) \tau_x - a_x (\hat{\E}[Y_i | D_i=0, X_i]  + \pi_x \tau_x) \\
    &= \sum_x P(X=x) \pi_x (1 - \pi_x) \tau_x - a_x \hat{\E}[Y_i|X_i=x]\\
\end{align}
Using this, we can write the weighted representation of the regression coefficient in terms of $a_x$:
\begin{align}
    \hat{\tau}_{reg} = \frac{\sum_x P(X=x) \pi_x (1 - \pi_x) \tau_x - a_x \hat{\E}[Y_i|X_i=x]}{\sum_x P(X=x) (\pi_x(1 - \pi_x - a_x) + (\pi_x + a_x)(\pi_x + a_x - \pi_x))}
\end{align}

\subsection{Equivalency of impute and interact} \label{equiv}

We can show that the regression imputation estimate is equivalent to the Lin/interacted regression estimate by showing that the minimization problems solved by the two methods are equivalent.

We can start with the Lin estimate. We want to estimate $\hat{\beta}$ in the equation:

$$\E[Y|D, X] = \hat{\beta}_0 + \hat{\beta}_1 D + \hat{\beta}_2 X + \hat{\beta}_3 DX$$

We do this by minimizing the least squares error, in other words, $\hat{\beta}$ is the solution to the minimization problem:

$$\min_{\beta} \sum_{i=1}^n (Y_i - (\beta_0 + \beta_1 D_i + \beta_2 X_i + \beta_3 D_iX_i))^2$$

Now let's look at the regressions involved in the imputation estimate. The imputation estimate is $\hat{\theta} = \frac{1}{N} \sum_{i=1}^N \hat{Y}_i(1) - \hat{Y}_i(0) = \frac{1}{N} \sum_{i=1}^N \gamma_0 + \gamma_1 X - (\alpha_0 + \alpha_1 X)$. Here, $\gamma$ and $\alpha$ are found by solving the following minimization problems:

\begin{align}
    &\min_{\gamma} \sum_{D_i=1} (Y_i - (\gamma_0 + \gamma_1 X_i))^2 \\
    &\min_{\alpha} \sum_{D_i=0} (Y_i - (\alpha_0 + \alpha_1 X_i))^2 \\
\end{align}

This is equivalent to solving the single minimization problem:
\begin{align}
&=\min_{\gamma, \alpha} \sum_{D_i=1} (Y_i - (\gamma_0 + \gamma_1 X_i))^2 + \sum_{D_i=0} (Y_i - (\alpha_0 + \alpha_1 X_i))^2 \\
&=\min_{\gamma, \alpha} \sum_{i=1}^n D_i(Y_i - (\gamma_0 + \gamma_1 X_i))^2 + (1 - D_i)(Y_i - (\alpha_0 + \alpha_1 X_i))^2 \\
&=\min_{\gamma, \alpha} \sum_{i=1}^n D_i(Y_i^2 - 2Y_i(\gamma_0 + \gamma_1 X_i) + (\gamma_0 +\gamma_1 X_i)^2) + (1 - D_i)(Y_i^2 - 2 Y_i (\alpha_0 \\
&+ \alpha_1 X_i) + (\alpha_0 + \alpha_1 X_i)^2) \nonumber \\
&=\min_{\gamma, \alpha} \sum_{i=1}^n Y_i^2 - 2Y_i(\alpha_0 + \alpha_1 X_i + (\gamma_0 - \alpha_0)D_i + (\gamma_1 - \alpha_1) D_i X_i) + (\alpha_0 + \alpha_1 X_i + (\gamma_0 - \alpha_0)D_i \\
&+ (\gamma_1 - \alpha_1) D_i X_i)^2 \nonumber \\
&=\min_{\gamma, \alpha} \sum_{i=1}^n (Y_i - (\alpha_0 + \alpha_1 X_i + (\gamma_0 - \alpha_0)D_i + (\gamma_1 - \alpha_1) D_i X_i))^2 \\
&=\min_{\beta} \sum_{i=1}^n (Y_i - (\beta_0 + \beta_1 D_i + \beta_2 X_i + \beta_3 D_iX_i))^2
\end{align}
where $\beta_0 = \alpha_0$, $\beta_1 = \gamma_0 - \alpha_0$, $\beta_2 = \alpha_1$, and $\beta_3 = \gamma_1 - \alpha_1$

\subsection{Unbiasedness of weighted DiM using mean balancing weights}
\label{meanbal-unbiased}

Assume the potential outcomes are separately linear in $X$: 
\begin{align}
    \E[Y_i(1) | X_i] &= \gamma X_i \\
    \E[Y_i(0) | X_i] &= \alpha X_i
\end{align}

Then we can show that the ATE estimate from using mean balancing weights is unbiased:
\begin{align}
    \E\left[\hat{\tau}_{meanbal}\right] &= \E\left[\sum_{D_i = 1} Y_i w_i - \sum_{D_i = 0} Y_i w_i\right] \\
    &=  \sum_{D_i=1} \E[Y_i w_i | D_i = 1] -  \sum_{D_i=0} \E[Y_i w_i | D_i = 0] \\
    &=  \sum_{D_i=1} \E[Y_i(1) w_i | D_i = 1] -  \sum_{D_i=0} \E[Y_i(0) w_i | D_i = 0] \\
    &=  \sum_{D_i=1} \E[(\gamma X_i +\epsilon_i) w_i | D_i = 1] -  \sum_{D_i=0} \E[(\alpha X_i + \epsilon_i) w_i | D_i = 0] \\
    &= \gamma \sum_{D_i=1} \E[{X}_i w_i | D_i = 1] - \alpha \sum_{D_i=0} \E[{X}_i w_i | D_i = 0] \\
    &= \gamma  \E \left[ \sum_{D_i=1} {X}_i w_i  \right] - \alpha \E\left[\sum_{D_i=0} {X}_i w_i \right] \\
    &= \gamma  \E \left[ \bar{{X}} \right] - \alpha  \E\left[ \bar{{X}} \right] \\
    &= \E[\overline{Y_i(1)}] - \E[\overline{Y_i(0)}] \\
    &= \E[Y_i(1) - Y_i(0)]
\end{align}

where $\overline{{X}} = \frac{1}{N} \sum_{i=1}^N X_i$ is the sample mean of the covariates.

\subsection{Unbiasedness of impute and interact}

We can show that both the regression imputation method and the interacted regression are unbiased for the ATE. We start with regression imputation. We have one model for $Y$ among the treated, $\hat{\mu}_1$ , and one among the controls, $\hat{\mu}_0$. 

Then,
\begin{align}
\hat{\tau}_{imp} &= \frac{1}{n} \left(\sum_{D_i=1} (Y_i - \hat{\mu}_0(X_i)) + \sum_{D_i = 0} (\hat{\mu}_1(X_i) - Y_i )\right) \\
&= \frac{1}{n} \sum_i (\hat{\mu}_{1}(X_i) - \hat{\mu_{0}}(X_i)) \\
&= \sum_{x \in \mathcal{X}} \frac{\mathds{1}_{\{X = x\}}}{n}\left( \hat{\mu}_1(x) - \hat{\mu}_0(x)\right) \\
&= \sum_{x \in \mathcal{X}} \widehat{P}(X=x) \left(\hat{\mu}_1(x) - \hat{\mu}_0(x)\right) \\
&= \sum_{x \in \mathcal{X}} \widehat{P}(X=x) DIM_x
\end{align}

We see that the imputation approach weights each stratum by the probability of inclusion in that stratum -- these weights result in an estimate that is unbiased for the ATE.

\end{document}